\documentclass[12pt]{iopart}

\usepackage{amssymb}
\usepackage[dvips]{graphics,graphicx}
\usepackage{psfrag}

\newcommand{\ed}{\hbox{\rlap{$\sqcap$}$\sqcup$}}
\newtheorem{defi}{Definition}[section]
\newtheorem{teo}{Theorem}[section]

\newtheorem{coro}{Corollary}[section]

\newenvironment{demo}{{\em Proof.\mbox{ }}}{{$\Box$}}
\newcommand{\bdefi}{\begin{defi}}
\newcommand{\edefi}{\end{defi}}
\newcommand{\ben}{\begin{enumerate}}
\newcommand{\eenu}{\end{enumerate}}
\newtheorem{Lemma}{Lemma}[section]
\newtheorem{definition}{Definition}[section]
 
\newtheorem{example}{Example}[section]

\begin{document}
\title[Combinatorial approach to generalized Bell and Stirling
numbers...]{Combinatorial approach to generalized Bell and Stirling
numbers and boson normal ordering problem}

\author{M A M\'{e}ndez$^{\sharp,\$}$, P Blasiak$^{*,\diamondsuit}$
and K A Penson$^{*}$}

\address
{$^\sharp$ IVIC, Lab. Anal. Mat.-Matem\'{a}ticas,\\
Carretera Panamericana, Km.11, Caracas, Venezuela}

\address
{$^\$$ UCV, Dpto. de Matem\'{a}ticas, Facultad de Ciencias\\ Caracas,
Venezuela}

\address
{$^*$ Laboratoire de Physique Th\'{e}orique de la Mati\`{e}re Condens\'{e}e\\
Universit\'{e} Pierre et Marie Curie, CNRS UMR 7600\\
Tour 24 - 2\`{e}me \'{e}t., 4 pl. Jussieu, F 75252 Paris Cedex 05, France}

\address
{$^\diamondsuit$ H.Niewodnicza\'nski Institute of
Nuclear Physics, Polish Academy of Sciences\\
ul. Eliasza-Radzikowskiego 152,  PL 31342 Krak\'{o}w, Poland}

\eads{\linebreak \mailto{mmendez@cauchy.ivic.ve},
\mailto{blasiak@lptl.jussieu.fr},
\mailto{penson@lptl.jussieu.fr}}

\begin{abstract}
We consider the numbers arising in the problem of normal ordering
of expressions in boson creation $a^\dag$ and annihilation $a$
operators ($[a,a^\dag]=1$). We treat a general form of a boson
string
$(a^\dag)^{r_n}a^{s_n}...(a^\dag)^{r_2}a^{s_2}(a^\dag)^{r_1}a^{s_1}$
which is shown to be associated with generalizations of Stirling
and Bell numbers. The recurrence relations and closed-form
expressions (Dobi\'nski-type formulas) are obtained for these
quantities by both algebraic and combinatorial methods. By
extensive use of methods of combinatorial analysis we prove the
equivalence of the aforementioned problem to the enumeration of
special families of graphs. This link provides a combinatorial
interpretation of the numbers arising in this normal ordering
problem.
\end{abstract}

\section{Introduction}

In this paper we consider a pair of boson creation $a^\dag$ and
annihilation $a$ operators satisfying the commutation relation
\begin{eqnarray}\label{HW}
[a,a^\dag]=1.
\end{eqnarray}
These operators play a fundamental role in the formalism of second
quantization in Quantum Mechanics and Quantum Field Theory (QFT)
\cite{Baez},\cite{Louisell},\cite{Peshkin}. Since the creation and
annihilation operators do not commute serious problems with their
ordering arise. A very convenient and well defined form of the
operators depending on $a$ and $a^\dag$ is the so called normally
ordered form \cite{Klauder}. An operator is said to be in a
normally ordered form if all creation operators stand to the left
of the annihilation operators. The most important application
field of the normal order is the QFT \cite{Peshkin}. For a recent
study of the interplay of the QFT, normal order and combinatorics
see Ref.\cite{JMP}. Procedure of normal ordering of the operator,
i.e. moving all the creation operators to the left with the use of
relation Eq.(\ref{HW}), is in general a nontrivial task. A first
example is ordering of the power of the number operator
\cite{Katriel}\cite{Katriel2000}:
\begin{eqnarray}\label{aa}
(a^\dag a)^n=\sum_{k=1}^n S(n,k) (a^\dag)^ka^k,
\end{eqnarray}
where $S(n,k)$ are the Stirling numbers of the second kind
\cite{Comtet} enumerating partitions of the set of $n$ elements
into $k$ nonempty subsets, and satisfying the following recurrence
relation $S(n+1,k)=S(n,k-1)+kS(n,k)$ with initial values
$S(n,n)=S(n,1)=1$.
\\
As the extension of this result we have considered operators in
the form $(a^\dag)^ra^s$ ($r$, $s$ -positive integers, $r\geq s$),
for which a normally ordered form is given by
\begin{eqnarray}\label{Srs}
[(a^\dag)^ra^s]^n=(a^\dag)^{n(r-s)}\sum_{k=s}^{ns} S_{r,s}(n,k)
(a^\dag)^ka^k,
\end{eqnarray}
where $S_{r,s}(n,k)$ are generalized Stirling numbers
\cite{Blasiak1},\cite{Blasiak2},\cite{Blasiak3},\cite{Lang},\cite{Schork},\cite{Thesis}.
This kind of formulas allow one to write the exponentials
$e^{\lambda (a^\dag)^ra^s}$ in the normally ordered form and then
easily calculate the coherent state expectation values which are
of importance e.g. in Quantum Optics \cite{Klauder}. The clue of
these calculations is the knowledge of the properties of the
numbers $S_{r,s}(n,k)$. As they are of a combinatorial origin, the
recurrence relations, Dobi\'{n}ski-type formulas, closed-form
expresions and generating functions were extensively studied
\cite{Blasiak2}.
\\
In the following we further extend these results to normal
ordering of a general {\em boson string} in the form
$(a^{\dagger})^{r_n}a^{s_n}\dots(a^{\dagger})^{r_2}a^{s_2}(a^{\dagger})^{r_1}a^{s_1}$,
by establishing a link to special structures in enumerative
combinatorics. This in turn gives us the rigorous demonstration of
the properties of the generalized Stirling and Bell numbers
arising in this problem. The construction of the graphs (the so
called 'bugs') associated with these numbers provides a graphical
interpretation of the normal ordering procedure.

\section{Generalized Bell and Stirling numbers}

In this section we define the generalization of ordinary Bell and
Stirling numbers which arise in the solution of the normal
ordering problem for a boson string. Given two sequences of
positive integers ${\mathbf r}=(r_1,r_2,\dots,r_n)$ and ${\mathbf
s}=(s_1,s_2,\dots,s_n)$ we let $S_{{\mathbf r},{\mathbf s}}(k)$ be
the positive integers appearing in the expansion
\begin{eqnarray}\label{S}
(a^{\dagger})^{r_n}a^{s_n}\dots(a^{\dagger})^{r_2}a^{s_2}(a^{\dagger})^{r_1}a^{s_1}=
(a^{\dagger})^{d_n}\sum_{k=s_1}^{s_1+s_2+\dots+s_n}S_{{\mathbf
r},{\mathbf s}}(k)(a^{\dagger})^ka^k,
\end{eqnarray}
where $d_n=\sum_{i=1}^n(r_i-s_i)$, which we assume here to be
non-negative. We observe that the whole theory can be carried
through for $d_n$ negative, at the cost of minor adaptations,
which however do not change the numbers involved. Note that the
r.h.s. of Eq.(\ref{S}) is already normally ordered.
\\
We call $S_{{\mathbf r},{\mathbf s}}(k)$ the generalized Stirling
numbers of the second kind. The generalized Bell number is defined
as the sum
\begin{eqnarray}
B_{{\mathbf r},{\mathbf s}}=
\sum_{k=s_1}^{s_1+s_2+\dots+s_n}S_{{\mathbf r},{\mathbf s}}(k).
\end{eqnarray}
In this notation the generalized Stirling numbers defined in
Eq.(\ref{Srs}) correspond to a {\em uniform} case ${\mathbf
r}=\overbrace{(r,r,\dots,r)}^{n}$ and ${\mathbf
s}=\overbrace{(s,s_,\dots,s)}^{n}$.
\\
We introduce the notation ${\mathbf r}\ \uplus\
r_{n+1}=(r_1,r_2,\dots,r_n,r_{n+1})$ and ${\mathbf s}\ \uplus\
s_{n+1}=(s_1,s_2,\dots,s_n,s_{n+1})$ and state the recurrence
relation satisfied by generalized Stirling numbers $S_{{\mathbf
r},{\mathbf s}}(k)$
\begin{eqnarray}\label{rec}
S_{{\mathbf r}\uplus r_{n+1},{\mathbf s}\uplus
s_{n+1}}(k)=\sum_{j=0}^{s_{n+1}}
\left(\!\!\!\begin{array}{c}s_{n+1}\\j\end{array}\!\!\!\right)(d_n+k-j)_{s_{n+1}-j} S_{{\mathbf r},{\mathbf
s}}(k-j),
\end{eqnarray}
where $(l)_p=l\cdot(l-1)\cdot...\cdot(l-p+1)$ is the falling
factorial.
\newline
One can give the derivation of Eq.(\ref{rec}) by induction using
the following consequence of Eq.(\ref{HW}) (see the proof in
\cite{Louisell}):
\begin{eqnarray}
a^k(a^\dag)^l=\sum_{p=0}^k\left(\!\!\begin{array}{c}k\\p\end{array}\!\!\right)(l)_p(a^\dag)^{l-p}a^{k-p}.
\end{eqnarray}
The full details of this approach can be consulted in
\cite{Thesis}.
\\
Observe that the problem stated above can also be formulated in
terms of the multiplication $X$ and derivative $D$ operators as
they satisfy $[D,X]=1$. The representation of boson commutation
relation with the $D$ and $X$ operators resembles the Bargmann
representation \cite{Klauder}, used in connection with coherent
states. (Here we do not enter into that framework, with all the intricacies of the
scalar product, hermiticity etc., as in our context only the
algebraic properties matter.) Then Eq.(\ref{S}) can be rewritten
as:
\begin{eqnarray}\label{DX}
X^{r_n}D^{s_n}\dots X^{r_2}D^{s_2}X^{r_1}D^{s_1}=
X^{d_n}\sum_{k=s_1}^{s_1+s_2+\dots+s_n}S_{{\mathbf r},{\mathbf
s}}(k)X^kD^k.
\end{eqnarray}
Acting with both sides of Eq.(\ref{DX}) on the
exponential function $e^{x}$ we get the identity
\begin{eqnarray}\label{Defbell}X^{r_n}D^{s_n}\dots X^{r_2}D^{s_2}X^{r_1}D^{s_1}e^x=x^{d_n}e^xB_{{\mathbf r},{\mathbf s}}(x)
\end{eqnarray}
where
\begin{eqnarray}\label{BellPol}
B_{{\mathbf r},{\mathbf
s}}(x)=\sum_{k=s_1}^{s_1+s_2+\dots+s_n}S_{{\mathbf r},{\mathbf
s}}(k)x^k
\end{eqnarray}
is the so called generalized Bell polynomial. Observe that the
order of the so defined generalized Bell polynomial does not
depend on ${\mathbf r}$. Eq.(\ref{Defbell}) gives the
formula
\begin{eqnarray}
X^{d_{n+1}}e^xB_{{\mathbf r}\uplus r_{n+1},{\mathbf s}\uplus
s_{n+1}}(x)&=&X^{r_{n+1}}D^{s_{n+1}}e^x
x^{d_n}B_{{\mathbf r},{\mathbf s}}(x).
\end{eqnarray}
Using the well known commutation rule (equivalent to the Leibniz rule) $D^ne^xf(x)=e^x(D+I)^nf(x)$
we get the recursive formula
\begin{eqnarray}\label{rec1}
B_{{\mathbf r}\uplus r_{n+1},{\mathbf s}\uplus
s_{n+1}}(x)=X^{s_{n+1}-d_n}(D+I)^{s_{n+1}} X^{d_n}B_{{\mathbf
r},{\mathbf s}}(x)
\end{eqnarray}
By taking coefficients of $x^k$ on both sides of Eq.(\ref{rec1})
we obtain the recurrence relation for the generalized Stirling
numbers of Eq.(\ref{rec}).
\\
Observe that the action of the l.h.s.
of Eq.(\ref{DX}) on $e^x$ may be calculated explicitly. To this end one
first evaluates it on the monomial $x^n$ yielding
$\left[\prod_{j=1}^n(d_{j-1}+n)_{s_j}\right]x^{n+d_n}$ which in turn
easily gives the result of the action on the exponential function
$e^{x}$.
\newline
With this observation, together with Eq.(\ref{Defbell}) we arrive
at the extended Dobi\'{n}ski-type relation
 for generalized Bell polynomials
\begin{eqnarray}\label{dob}
B_{{\mathbf r},{\mathbf s}}(x)=e^{-x}\sum_{m=s_1}^\infty
\left[\prod_{j=1}^n(m+d_{j-1})_{s_j}\right]\frac{x^m}{m!},
\end{eqnarray}
which by Cauchy's multiplication of series yields the expression
for $S_{{\mathbf r},{\mathbf s}}(k)$:
\begin{equation}\label{s}
S_{{\mathbf r},{\mathbf
s}}(k)=\frac{1}{k!}\sum_{m=0}^k\left(\!\!\!\begin{array}{c}k\\m\end{array}\!\!\!\right)(-1)^{k-m}.
\prod_{j=1}^n(m+d_{j-1})_{s_j}
\end{equation}
An alternative, very similar demonstration of the above results
can be carried through with the use of coherent states. These are
defined for complex $z$, as
$|z\rangle=e^{-|z|^2/2}\sum_{n=0}^\infty
\frac{z^n}{\sqrt{n!}}|n\rangle$, where $a^\dag a|n\rangle
=n|n\rangle$, $a|z\rangle=z|z\rangle$ and $\langle
n|n'\rangle=\delta_{n,n'}$ \cite{Klauder}. The $|n\rangle$'s are
called the number states. The coherent state matrix element of
Eq.(\ref{S}) establishes a link to generalized Bell polynomials of
Eq.(\ref{BellPol}):
\begin{eqnarray}
\langle
z|(a^{\dagger})^{r_n}a^{s_n}\dots(a^{\dagger})^{r_2}a^{s_2}(a^{\dagger})^{r_1}a^{s_1}|z\rangle
=(z^*)^{d_n}B_{{\mathbf r},{\mathbf s}}(|z|^2),
\end{eqnarray}
which after some algebra, provides an equivalent derivation of
Eqs.(\ref{dob}) and (\ref{s}). The first instance where the
relation between the coherent state matrix elements and the Bell
polynomials appears is Ref.\cite{Katriel2000}, again for the
generic case of Eq.(\ref{aa}), for which conventional Bell
polynomials are obtained.
\\
We shall proceed now to give a combinatorial interpretation of the
above results. The essence of subsequent paragraphs will be a
graph-theoretical description of the problem. We define the
structures (graphs) that are counted by the generalized Bell and
Stirling numbers and then give a thorough combinatorial derivation
of the recurrence relations, Dobi\'{n}ski-type formulas and other
closed-form expressions. The Eqs.(\ref{rec}), (\ref{dob}) and (\ref{s}) will
emerge from purely combinatorial considerations and this will
permit the bijective identification of algebraic and combinatorial
structures.

\section{Bugs, colonies, settlements and recurrence relations}

We introduce now a number of tools needed to describe the problem
in the graph-theoretical language.
\begin{figure}[ht]
  \centering
   \includegraphics[height=3.2cm]{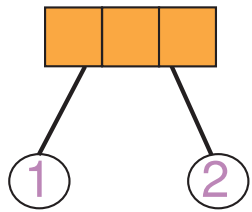}
   \caption{A $(3,2)$-bug}
   \label{bug}
 \end{figure}

\begin{defi}
\rm{A {\em bug} of {\em type} $(r,s)$ consists of a body and $s$
legs. The body is formed by $r$ linearly ordered empty cells. Each
foot of the $s$ legs is labelled with one number from an integer
segment $(m,m+s]:=\{m+1,m+2,\dots,m+s\}$, see Fig.\ref{bug}.}
\end{defi}

\begin{figure}[ht]
  \centering
   \includegraphics[height=7cm]{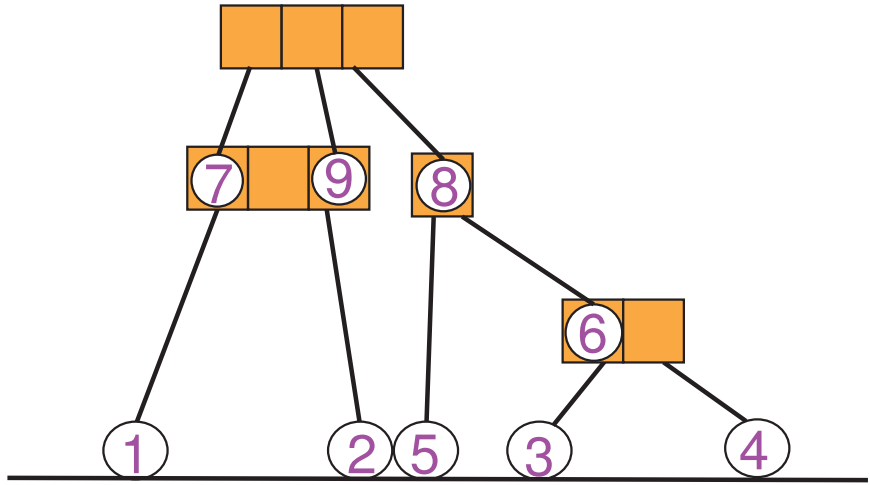}
   \caption{A colony of type (3,2,1,3;2,2,2,3) and $5$ free legs.}
   \label{colony}
 \end{figure}

\begin{defi}\label{first}
\rm{Consider a set of $n$ bugs, the first one of type $(r_1,s_1)$
and feet-labelled with labels in $(0,s_1]$, the second of type
$(r_2,s_2)$ with labels in $(s_1,s_1+s_2]$ and so on. A {\em
colony} is one of the possible ways of organizing the bugs using
the following procedure. The first bug has to stand over the
ground. Once the $(j-1)$th bug is placed, the $j$th is placed by
putting some (or none) of its $s_j$ feet in the ground and each
one of the rest in one of the empty cells of the bodies of the
preceding bugs, see Fig.\ref{colony}. The pair of sequences
$({\mathbf r},{\mathbf s})$, ${\mathbf r}=(r_1,r_2,\dots,r_n)$,
${\mathbf s}=(s_1,s_2,\dots,s_n)$, carrying the information about
the types of the bugs is called the {\em type} of the colony. The
legs of the colony standing on the ground are called {\em free}.
\\
Assume now that there is a set of $m$ empty cells in the ground.
An $m$-{\em settlement} is a colony where each one of the feet
corresponding to the free legs is placed in one of the ground
cells. A {\em surjective settlement} is one where all the ground
cells are occupied. The type of a settlement is defined to be the
type of the subjacent colony.}\end{defi}
The main theorem of interest for us is:

\begin{teo}\label{principal}
The Stirling number $S_{{\mathbf r},{\mathbf s}}(k)$, $s_1\leq
k\leq s_1+s_2+\dots+s_n$, counts the number of colonies of type
$({\mathbf r},{\mathbf s})$ having exactly $k$ free legs. The Bell
number $B_{{\mathbf r},{\mathbf s}}$ counts the number of colonies
of type $({\mathbf r},{\mathbf s})$.
\end{teo}
Before proving it, we state the following
\begin{Lemma}\label{cells}
{\rm A colony of type $({\mathbf r},{\mathbf s})$ and with  $k$
free legs has exactly $d_n+k$ empty cells.}
\end{Lemma}
\begin{demo}
The total number of cells of the colony is equal to $\sum_{i=1}^n
r_i$. The number of occupied cells is equal to the total number of
legs minus the number of free legs ($\sum_{i=1}^n s_i -k$).
\end{demo}\vspace{0.3cm}\\
Now we are ready to prove the Theorem \ref{principal}.\\
\begin{demo}  Denote by
$C_{\mathbf{r},\mathbf{s}}(k)$ the number of colonies of type
$(\mathbf{r},\mathbf{s})$ with exactly $k$ free legs. Since
$C_{(r_1;s_1)}(k)=S_{(r_1;s_1)}(k)=\delta(s_1,k)$ it is enough to
prove that the numbers $C_{\mathbf{r},\mathbf{s}}(k)$ satisfy the
same recursion as the generalized Stirling numbers of Eq.(\ref{rec}).
\begin{equation}
C_{{\mathbf r}\uplus r_{n+1},{\mathbf s}\uplus
s_{n+1}}(k)=\sum_{j=0}^{s_{n+1}}
\left(\!\!\!\begin{array}{c}s_{n+1}\\j\end{array}\!\!\!\right)(d_n+k-j)_{s_{n+1}-j}\ 
C_{{\mathbf r},{\mathbf s}}(k-j) \label{recursion}
\end{equation}
The l.h.s. is the number of colonies of type $({\mathbf r}\uplus
r_{n+1},{\mathbf s}\uplus s_{n+1})$ having exactly $k$ free legs.
We claim that in the right hand side the expression
$$\left(\!\!\!\begin{array}{c}s_{n+1}\\j\end{array}\!\!\!\right)(d_n+k-j)_{s_{n+1}-j}\ 
C_{{\mathbf r},{\mathbf s}}(k-j)$$ gives the number of such
colonies where the $(n+1)$th bug has exactly $j$ free legs.
Obviously, this would prove the identity.
 We now prove our claim. In order to get a colony with $k$ free
legs the $(n+1)$th bug has to be placed in a colony of type
$({\mathbf r},{\mathbf s})$ and $k-j$ free legs. $C_{{\mathbf
r},{\mathbf s}}(k-j)$ is the number of such colonies. We choose
the free legs of the $(n+1)$th bug in
$\left(\!\!\!\begin{array}{c}s_{n+1}\\j\end{array}\!\!\!\right)$
ways. Since by proposition (\ref{cells}) there are $d_n+k-j$ empty
cells in the $n$-bugs colony, $(d_n +k-j)_{s_{n+1}-j}$ gives the
number of ways of distributing the rest of the feet of the
$(n+1)$th bug into the empty cells.
\end{demo}
\begin{figure}[ht]
  \centering
   \includegraphics[height=7cm]{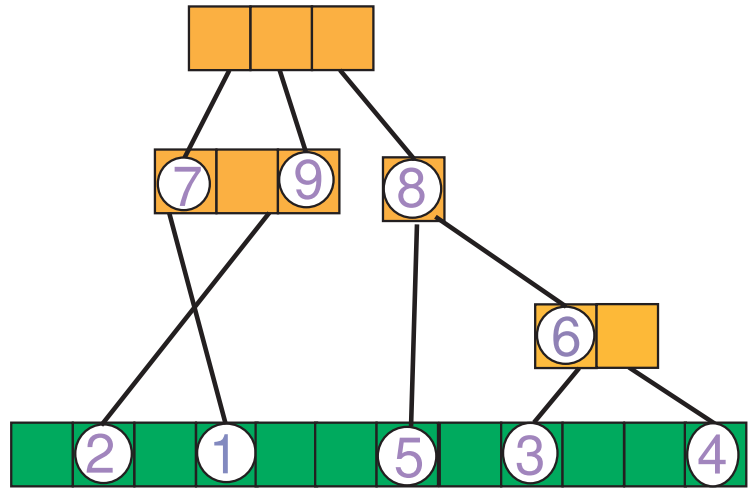}
   \caption{A $12$-settlement of the colony in Fig.(\ref{colony})}
   \label{settlement}
 \end{figure}
\\
In the next section will follow a number of propositions
clarifying the properties of structures in question.

\section{Counting settlements and Dobi\'{n}ski-type Relations}
We shall count now the number of $m$-settlements which will provide the
link with Eqs.(\ref{dob}) and (\ref{s}) viewed from the combinatorial perspective.
\begin{teo}
Let $p(m,{\mathbf r},{\mathbf s})$ be the number of
$m$-settlements of type $({\mathbf r},{\mathbf s})$. We have
\begin{equation}
p(m,{\mathbf r},{\mathbf s})=\prod_{j=1}^n(m+d_{j-1})_{s_j}.
\end{equation}
\end{teo}
\begin{demo}
There are $(m)_{s_1}$ ways of placing the feet of the first bug
into the $m$ ground cells. After placing the $(j-1)$th bug there
are $m+d_{j-1}$ empty cells available (the previously placed bugs
have provided $\sum_{i=1}^{j-1}r_i$ empty cells and occupied
$\sum_{i=1}^{j-1}s_i$ cells). Then, there are $(m+d_{j-1})_{s_j}$
ways of placing the $s_j$ feet of the $j$th bug.
\end{demo}
\begin{coro}
{\rm We have the polynomial identity}
\begin{equation}
\prod_{j=1}^n(x+d_{j-1})_{s_j}=\sum_{k=s_1}^{s_1+s_2+\dots+s_n}S_{{\mathbf
r},{\mathbf s}}(k)(x)_k.
\end{equation}
\end{coro}
\begin{demo}
By the previous theorem, for an integer value of $x$ the l.h.s.
 counts the number of $x$-settlements of type $({\mathbf
r},{\mathbf s})$. $S_{{\mathbf r},{\mathbf s}}(k)(x)_k$ counts the
number of ways of settling a colony of type $({\mathbf r},{\mathbf
s})$ with $k$ free legs in $x$ ground cells. Then, the r.h.s. is
another way of counting $x$-settlements.
\end{demo}
\vspace{0.2cm}\\
The exponential generating function of the surjective settlements
is equal to the polynomials $$B_{{\mathbf r},{\mathbf
s}}(x)=\sum_{k=s_1}^{s_1+s_2+\dots+s_n}S_{{\mathbf r},{\mathbf
s}}(k)k!\frac{x^k}{k!}=
\sum_{k=s_1}^{s_1+s_2+\dots+s_n}S_{{\mathbf r},{\mathbf
s}}(k)x^k.$$
\begin{coro}{\rm (Extended Dobi\'{n}ski-type
relations)
\\
We have the identity}
\begin{equation}\label{Dobinski}
B_{{\mathbf r},{\mathbf s}}(x)e^{x}=\sum_{m=s_1}^\infty
\prod_{j=1}^n(m+d_{j-1})_{s_j}\frac{x^m}{m!}.
\end{equation}
\end{coro}
\begin{demo}Taking the coefficient of $\frac{x^m}{m!}$ of the left
hand side we obtain $$\sum_{k=0}^m
\left(\!\!\begin{array}{c}m\\k\end{array}\!\!\right)S_{{\mathbf r},{\mathbf
s}}(k)k!=\sum_{k=0}^m S_{{\mathbf r},{\mathbf s}}(k)(m)_k.$$ By
the previous corollary it is equal to the coefficient of
$\frac{x^m}{m!}$ in the right hand side.
\end{demo}
\vspace{0.2cm}\\
From Eq.(\ref{Dobinski}) we obtain
\begin{equation}
\label{Dobinski1} B_{{\mathbf r},{\mathbf
s}}(x)=e^{-x}\sum_{m=s_1}^\infty
\prod_{j=1}^n(m+d_{j-1})_{s_j}\frac{x^m}{m!}
\end{equation}
and
\begin{equation}\label{Dobinski2} B_{{\mathbf
r},{\mathbf s}}\equiv B_{{\mathbf r},{\mathbf
s}}(1)=e^{-1}\sum_{m=s_1}^\infty\frac{1}{m!}
\prod_{j=1}^n(m+d_{j-1})_{s_j}\ .
\end{equation}
Taking the coefficient of $\frac{x^k}{k!}$ on both sides of
equation (\ref{Dobinski1}) we obtain the formula for the
generalized Stirling numbers
\begin{equation}\label{ss}
S_{{\mathbf r},{\mathbf
s}}(k)=\frac{1}{k!}\sum_{m=0}^k\left(\!\!\begin{array}{c}k\\m\end{array}\!\!\right)(-1)^{k-m}
\prod_{j=1}^n(m+d_{j-1})_{s_j}.
\end{equation}
Evidently Eq.(\ref{Dobinski1}) is identical to Eq.(\ref{dob}) and
so are Eqs.(\ref{ss}) and (\ref{s}). This emphasizes again the already stated
bijective correspondence between algebraic and combinatorial
structures.

\section{Uniform colonies and settlements}
\begin{figure}[ht]
  \centering
   \includegraphics[height=4cm]{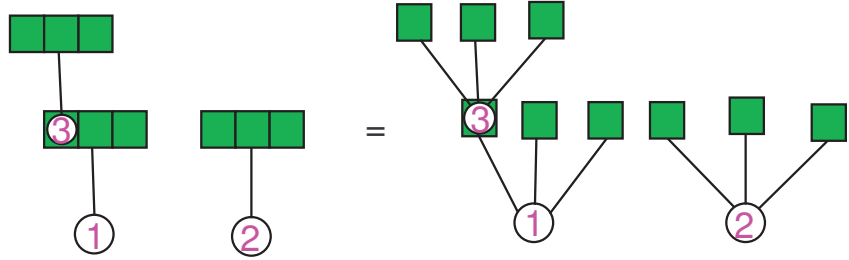}
   \caption{A uniform colony of type $(3,1)^3$ and the corresponding forest
 of increasing trees.}
   \label{forest}
 \end{figure}

A colony or a settlement with all the bugs of the same type is
called {\em uniform}. A uniform colony with $n$ bugs of type
$(r,s)$ is called a colony of type $(r,s)^n$. Following the
notation of \cite {Blasiak2} the corresponding Stirling and Bell
numbers, enumerating uniform colonies of type $(r,s)^n$, are
denoted respectively by $S_{r,s}(n,k)$ and $B_{r,s}(n).$ Clearly,
for ${\mathbf r}=\overbrace{(r,r,\dots,r)}^{n}$ and ${\mathbf
s}=\overbrace{(s,s_,\dots,s)}^{n}$, $S_{r,s}(n,k)=S_{{\mathbf
r},{\mathbf s}}(k)$ and $B_{r,s}(n)=B_{{\mathbf r},{\mathbf s}}$.
The recursive formula, Dobi\'{n}ski-type relations and its
consequences appearing here are natural extensions of those
investigated in \cite{Blasiak2}.
\begin{figure}[ht]
  \centering
   \includegraphics[height=3.5cm]{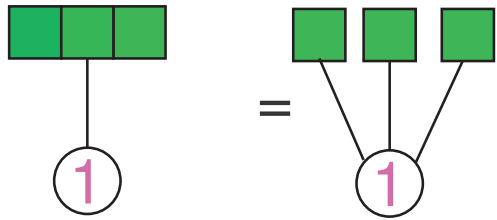}
   \caption{A $(3,1)$-bug and corresponding small planar
 tree.}
   \label{bush}
 \end{figure}
\\
The case $s=1$ can be mapped into trees and forests. An
$(r,1)$-bug can be identified with a planar tree, i.e. a tree
where the leaves are all connected to the root and linearly
ordered (see Fig.\ref{bush}). An increasing tree is one where the
internal vertices are labelled with labels in a totally ordered
set and the labels increase on any path from the root to any
internal vertex. The uniform colonies with $s=1$  corresponds to
forests of increasing $r$-ary planar trees. The free legs are the
roots of the trees (see Fig.\ref{forest}). For $r=1,$ there is
only one $1$-ary increasing tree for each $n$. Then
 $B_{1,1}(n)=B(n)$ is the ordinary Bell number.
\\
 The exponential generating function of the $r$-ary planar
 increasing trees $T_r(x)$ satisfy the differential equation (see
 \cite{B-L-L}, Chap. 5)
 $y'=(y)^r$. From this we obtain $T_r(x)=\sqrt[1-r]{1-(r-1)x},$ for
 $r>1$. The generalized Bell number
 $B_{r,1}(n)$ counts the number of $r$-forests with $n$ internal vertices.
 By the exponential formula we obtain
\begin{eqnarray}
\sum_{n=0}^{\infty}B_{r,1}(n)\frac{x^n}{n!}=\exp\{{\sqrt[1-r]{1-(r-1)x}-1}\}.
\end{eqnarray}
We quote the explicit expression \cite{Blasiak2}:
\begin{eqnarray}\label{B}
B_{r,1}(n)=\frac{(r-1)^{n-1}}{e}\sum_{k=1}^\infty
\frac{\Gamma(n+\frac{k}{r-1})} {\Gamma(1+\frac{k}{r-1})(k-1)!}.
\end{eqnarray}
In a subsequent publication we shall demonstrate that the
summation formulas of the type Eq.(\ref{B}) can be also obtained
for many other strings describing the uniform case.

\section{Conclusions}

We have obtained analytic expressions and combinatorial
interpretation of the integers generalizing conventional Bell and
Stirling numbers, arising in the normal ordering of a boson
string. All of their properties can be interpreted in terms of
graph-theoretical language. The proof of the main result may also
be obtained with the use of combinatorial theory of species
\cite{B-L-L},\cite{Joy},\cite{Flajolet}. The results constitute an
application of combinatorial analysis which produces the solution
of quantum mechanical problem of normal ordering. For alternative
interpretations of the numbers investigated in this work see
Refs.\cite{QTS3} and \cite{Varvak}. It is an outstanding problem
how to extend the key results of this work to the boson
$q$-analogs. In this respect the
Refs.\cite{Ehrenborg},\cite{KatrielKibler} and \cite{Czech} will
be of essential help.

\vspace{3mm}

\noindent We thank L. Haddad for important discussions.

\noindent PB wishes to thank the Polish Ministry of Scientific
Research and Information Technology for support under Grant no:
1P03B 051 26.

\end{document}